\documentstyle[twoside,fleqn,espcrc2]{article}

\newcommand{\AmS}{{\protect\the\textfont2
  A\kern-.1667em\lower.5ex\hbox{M}\kern-.125emS}}

\def\beq{\begin{equation}}
\def\eeq{\end{equation}}
\def\eq{\end{equation}}
\def\to{\rightarrow}

\def\EmissT{\not \! \!  E_{T}}

\def\Emiss{\not  \! \! E}

\newcommand{\newc}{\newcommand}

\newc{\eegg}{e^+e^-\gamma\gamma}
\newc{\mmgg}{\mu \mu \gamma\gamma}
\newc{\ttgg}{\tau \tau \gamma\gamma}

\newc{\leplep}{l^+l^-}
\newc{\llgg}{l^+l^- \gamma \gamma}
\newc{\lllgg}{l^+l^-l^{\prime \pm} \gamma \gamma}
\newc{\ljjgg}{l^{\pm}jj \gamma \gamma}
\newc{\lljjgg}{l^+l^-jj \gamma \gamma}

\newc{\eeggE}{ee \gamma \gamma + \EmissT}
\newc{\llggE}{l^+l^- \gamma \gamma + \EmissT}

\newc{\gag}{\gamma\gamma}
\newc{\lgg}{l^{\pm} \gamma \gamma}

\newc{\jjgg}{jj \gamma \gamma}
\newc{\jjjjgg}{4j \gamma \gamma}

\newc{\gsim}{\lower.7ex\hbox{$\;\stackrel{\textstyle>}{\sim}\;$}}
\newc{\lsim}{\lower.7ex\hbox{$\;\stackrel{\textstyle<}{\sim}\;$}}
\newc{\ie}{{\it i.e.}}
\newc{\etal}{{\it et al.}}
\newc{\eg}{{\it e.g.}}
\newc{\kev}{\hbox{\rm\,keV}}

\newc{\tbeta}{\tan\beta}
\newc{\uL}{{\tilde u_L}}
\newc{\uR}{{\tilde u_R}}
\newc{\cL}{{\tilde c_L}}
\newc{\cR}{{\tilde c_R}}
\newc{\tL}{{\tilde t_L}}
\newc{\tR}{{\tilde t_R}}
\newc{\dL}{{\tilde d_L}}
\newc{\dR}{{\tilde d_R}}
\newc{\sL}{{\tilde s_L}}
\newc{\sR}{{\tilde s_R}}
\newc{\bL}{{\tilde b_L}}
\newc{\bR}{{\tilde b_R}}
\newc{\eL}{{\tilde e_L}}
\newc{\eR}{{\tilde e_R}}
\newc{\mhp}{m_{H^\pm}}
\newc{\mhalf}{m_{1/2}}

\newc{\lR}{\tilde{l}_R}
\newc{\lL}{\tilde{l}_L}
\newc{\nL}{\tilde{\nu}_L}
\newc{\na}{\chi^0_1}
\newc{\nb}{\chi^0_2}
\newc{\nc}{\chi^0_3}
\newc{\nd}{\chi^0_4}
\newc{\ca}{\chi^{\pm}_1}
\newc{\cb}{\chi^{\pm}_2}
\newc{\capos}{\chi^{+}_1}
\newc{\caneg}{\chi^{-}_1}

\title{\hfill {\normalsize SLAC-PUB-7236} \\ 
\vspace{.2in}
Phenomenological Implications of 
Low Energy Supersymmetry Breaking}

\author{
Savas Dimopoulos,\address{Physics Department, Stanford University, 
Stanford, CA 94309}
\address{Theoretical Physics Division, CERN, CH-1211,
Geneva 23, Switzerland} 
Michael Dine,\address{Santa Cruz Institute for Particle
Physics, \\ University of California, Santa Cruz, CA 95064} 
Stuart Raby,\address{Physics Department, Ohio State University, 
Columbus, OH 43210}
Scott Thomas,\address{Stanford Linear Accelerator Center, 
Stanford, CA 94309}
\thanks{Talk presented by S. Thomas at the Fourth International Conference
on Supersymmetry (SUSY96), College Park, MD, 
May 29,1996.}
James D. Wells{\hbox{$^{\rm e}$}}
}

\begin{document}

\begin{abstract}
The experimental signatures for low energy supersymmetry breaking are
presented. 
The lightest standard model superpartner is unstable and 
decays to its partner plus a Goldstino, $G$. 
For a supersymmetry breaking scale below a few 1000 TeV
this decay can take place within a detector, leading
to very distinctive signatures. 
If a neutralino is the lightest standard model superpartner
it decays by $\chi_1^0 \to \gamma + G$, and if kinematically
accessible by $\chi_1^0 \to (Z^0, h^0, H^0, A^0) + G$.
These decays can give rise to displaced vertices.
Alternately, if a slepton is the lightest standard model superpartner
it decays by $\tilde{l} \to l + G$. 
This can be seen as a greater than minimum ionizing 
charged particle track, possibly with a kink to a minimum ionizing track. 
\end{abstract}

\maketitle

\section{Introduction}

If nature is supersymmetric, one of the most interesting questions 
to address experimentally is the scale and mechanism of supersymmetry
breaking. 
Most phenomenological studies of supersymmetric signals at high energy
colliders implicitly
assume that the messengers of supersymmetry breaking 
are gravitational strength interactions. 
The supersymmetry breaking scale in some hidden sector 
is then necessarily of order $10^{11}$ GeV.
If $R$-parity is conserved the lightest standard model superpartner
is the lightest supersymmetric particle (LSP) and is stable. 
If the LSP is electrically neutral, it escapes a detector leading
to the well known signal for supersymmetry of missing energy. 
It is possible however that the messenger scale for transmitting
supersymmetry breaking to the visible sector is anywhere between the
Planck scale and just above the electro-weak scale \cite{lsgauge,hsgauge}.
In this case the gravitino is the LSP, and the lightest standard 
model superpartner is the next to lightest supersymmetric particle
(NLSP).
The lightest standard model superpartner is unstable and decays
to its partner plus the Goldstino component of the gravitino \cite{fayet}.
For supersymmetry breaking scales below a few 1000 TeV this decay to 
the Goldstino can take place within a detector. 
This leads to very distinctive features for low scale supersymmetry 
breaking, including displaced photons, displaced charged particle or
$b$-jet vertices, or heavy charged sleptons possibly decaying to leptons 
within the detector \cite{ddrt}.
In this note we describe some of the model independent experimental signatures of 
low scale supersymmetry breaking. 

The role of the messenger sector is to couple the visible 
and supersymmetry breaking sectors.  
Integrating out the messenger sector gives rise to effective operators, which,
in the presence of supersymmetry breaking, lead to soft supersymmetry breaking
in the visible sector. 
If the messenger sector interactions are of gravitational strength,
the soft terms in the visible sector are logarithmically sensitive to 
ultraviolet physics all the way to the Planck or compactification
scale. 
In this case patterns within the 
soft terms might give an indirect window to Planck scale physics. 
However, the soft terms could then also be sensitive to 
some sector below the Planck scale which is responsible
for the flavor structure of the Yukawa interactions.
This generally leads to unacceptably large flavor changing neutral currents,
although elaborate flavor symmetries can be imposed to limit this.  
In contrast, if the messenger scale is well below the scale at which 
the Yukawa hierarchies are generated, the soft terms can be 
insensitive to the flavor sector, and naturally small flavor changing
neutral currents can result. 
The lack of flavor changing neutral currents is a significant advantage
of low scale supersymmetry breaking. 

In the next section the form the superpartner spectrum for the minimal
model of gauge-mediated supersymmetry breaking is reviewed. 
In section three the model independent experimental
signatures of supersymmetry breaking
at a low scale are presented. 

\section{Superpartner Spectrum with Gauge-Mediated Supersymmetry Breaking}

If supersymmetry is broken at a low scale it is in fact likely that 
the standard model gauge interactions play some role. 
This is because the standard model gauginos couple only through gauge 
interactions.
If standard model Higgs scalars received soft masses from non-gauge interactions,
the standard model gauginos would then be unacceptably lighter than 
the electro-weak scale.\footnote{The argument for light gauginos 
in the absence of gauge couplings within a low scale 
messenger sector 
only applies if the gauginos are elementary degrees of freedom in 
the ultraviolet, and would not apply if the gauginos where composite
or magnetic at or below the messenger scale.}

\begin{table}
\begin{center}
\begin{tabular}{cc}
\hline \hline
Particle &  Mass (GeV) \\
\hline
$\tilde{Q}_L, \tilde{Q}_R$  &  870, 835 \\
$\tilde{t}_1, \tilde{t}_2$  &  760, 860 \\
$A^0, H^0, H^{\pm}$ &   515-525 \\
$\chi_3^0, \chi_2^{\pm}, \chi_4^0$  &  415-440 \\
$\tilde{l}_L$   &  270 \\
${\chi_2^0}, {\chi_1^{\pm}}$  &  175 \\
$\tilde{l}_R$   &  140 \\
$h^0$  &  105 \\
${\chi_1^0}$  & 95 \\
$G$ & $1.5 \times 10^{-9}$ \\
\hline \hline
\end{tabular}
\caption{Typical spectrum with gauge-mediated supersymmetry
breaking for $\tan \beta = 3$, and a messenger scale of 
80 TeV.}
\end{center}
\end{table}

The standard model gauge interactions act as messengers
of supersymmetry breaking if 
fields within the supersymmetry breaking sector
transform under the standard model gauge group.
Integrating out these messenger sector fields then gives rise
to standard model gaugino masses at one-loop, and scalar
masses squared at two-loops \cite{lsgauge,hsgauge,dn}.
The scalar and gaugino masses are then very generally of the same 
order and go roughly as their gauge couplings squared. 
The $B$-ino and right handed sleptons gain masses through
$U(1)_Y$ interactions, and are therefore lightest. 
The $W$-ino's and left handed sleptons, transforming under $SU(2)_L$,
are somewhat heavier. 
The strongly interacting squarks and gluino are significantly
heavier than the electro-weak states. 

The dimensionful parameters within the Higgs sector,
$W = \mu H_u H_d$ and $V =  m_{12}^2 H_u H_d + h.c.$,
do not follow from the anzatz of gauge-mediated supersymmetry breaking,
and require
additional interactions which break 
$U(1)_{PQ}$ and $U(1)_{R-PQ}$ symmetries.
At present there is not a good model which gives rise these
Higgs sector masses without tuning parameters. 
The parameters $\mu$ and $m_{12}^2$ are therefore taken as free 
parameters in the minimal model, and can be eliminated in favor
of $\tan \beta$ and $m_Z$. 

Electroweak symmetry breaking results from the negative one-loop
correction to $m_{H_u}^2$ from stop-top loops 
due to the large top quark Yukawa coupling. 
Although this effect is formally three-loops, it is larger in magnitude 
than the electroweak contribution to $m_{H_u}^2$ due to the large squark masses.
Upon imposing electro-weak symmetry breaking, $\mu$ is typically found to be 
in the range $\mu \sim (1-2) m_{\tilde{l}_L}$ (depending on $\tan \beta$ and the 
messenger scale). 
This leads to a lightest neutralino, $\chi_1^0$, which is mostly
$B$-ino, and a lightest chargino, $\chi_1^{\pm}$, which is mostly $W$-ino. 
A typical spectrum at the low scale for a messenger
sector at 80 TeV transforming as ${\bf 5} + \bar{\bf 5}$ of 
$SU(5)$ is given in table 1. 

These general patterns within the spectrum should 
be considered model-dependent features of the 
minimal model of gauge-mediated supersymmetry
breaking.
The signatures for decay of the NLSP to its partner plus the Goldstino
discussed in the next section are however very model independent. 

\section{Detecting the Goldstino}

The spontaneous breaking of global supersymmetry leads to the 
existence of a massless Goldstone fermion, the Goldstino. 
In local supersymmetry the Goldstino becomes the longitudinal 
component of the gravitino. 
For supersymmetry breaking scales in the range discussed below,
the gravitino is essentially massless on the scale of 
accelerator experiments. 
The LSP is for all practical purposes 
the Goldstino, and the lightest standard model superpartner is the NLSP.

The lowest order couplings of the Goldstino are fixed by the 
supersymmetric Goldberger-Treiman low energy theorem to be 
given by \cite{fayet}
\beq
{\cal L} = - {1 \over F} j^{\mu \alpha} \partial_{\mu} G_{\alpha}
  ~ +~ h.c.
\eq
where $\sqrt{F}$ is the supersymmetry breaking scale, 
$j^{\mu \alpha}$ is the supercurrent, and $G_{\alpha}$ is the 
spin ${1 \over 2}$ Goldstino. 
Since the Goldstino acts like the supercharge, it transforms 
a superpartner into its partner. 
For $\sqrt{F}$ below a few 1000 TeV, such a decay of a superpartner to 
its partner plus the Goldstino can take place within a detector. 
Since the Goldstino couplings are suppressed compared to electroweak
and strong interactions, decay to the Goldstino is only relevant for the
lightest standard model superpartner (NLSP). 

The production of pairs of supersymmetric particles at a high
energy collider therefore takes place through standard model
couplings (assuming $R$-parity conservation). 
The produced states cascade to pairs of NLSP's. 
The quasi-stable NLSP's eventually decay to their partners plus
Goldstinos, which carry away missing energy. 
The specific signatures which arise from decay to the Goldstino
depend crucially on the quantum numbers of the NLSP, as discussed
in the following subsections. 

Associated production of Goldstinos in particle collisions has
been discussed in the past, but is completely irrelevant unless
$\sqrt{F}$ almost coincides with the electro-weak scale. 

\subsection{Neutralino NLSP}

It is possible that the lightest standard model superpartner 
is a neutralino, $\chi_1^0$.
It can decay through the gaugino components by
$\chi_1^0 \to \gamma + G$ and   
$\chi_1^0 \to Z^0 + G$, and through the Higgsino components by
$\chi_1^0 \to h^0 +  G$,
$\chi_1^0 \to H^0 +  G$, or 
$\chi_1^0 \to A^0 + G$ if kinematically accessible. 
These decays lead to very distinctive features if prompt on the scale of a detector. 

If $\chi_1^0$ is mostly gaugino, as in the example given in the 
previous section, it decays predominantly by $\chi_1^0 \to \gamma + G$. 
At an $e^+e^-$ collider the signature would be $e^+e^- \to \gamma \gamma + \Emiss$.
The standard model backgrounds are easily manageable for this 
signal \cite{ddrt,stump}.
Since $\chi_1^0$ is a Majorana particle, it can decay to both Goldstino 
helicities, giving an isotropic decay in the rest frame. 
The lab photon energy distribution is therefore flat and bounded by 
$ {1 \over 4} \sqrt{s} (1 - \beta) \leq E_{\gamma} \leq  {1 \over 4} \sqrt{s} (1 + \beta)$,
where $\beta$ is the $\na$ lab frame velocity. 
The end points of the photon spectrum give an important test that the final state
particles carrying the missing energy are essentially massless. 

At a hadron collider mostly gaugino $\na \na$ production is suppressed
by small couplings and large squark mass. 
Production of $\nb \ca$, $\capos \caneg$, and $\lR^+ \lR^-$ is however not 
suppressed. 
Cascade decays then lead to the final states 
$WZ \gag + \EmissT$ or $W l^+ l^- + \EmissT$, $WW \gag + \EmissT$, 
and $\llgg + \EmissT$, all at comparable rates \cite{dtw}.
For $m_{\na} \simeq 100$ GeV the total cross section in all such 
channels at the Tevatron can be up to 70 fb,
and could therefore be seen in current data. 
The existence of two hard $\gamma$'s in such events gives a distinctive
signature for low scale supersymmetry breaking. 
In addition, the backgrounds are much smaller than 
for standard supersymmetric signals \cite{ddrt,dtw,gordy},
with mis-identifications probably being the largest contamination of the signal. 
If the $\na$ decay length is long enough, timing information can also
be used to isolate the signal.

If such signatures were observed experimentally, one of the most important 
challenges would be to measure the distribution of finite path lengths for the
decaying $\na$'s, thereby giving a direct measurement of the supersymmetry breaking
scale. 
For $\sqrt{F}$ between roughly 100-1000 TeV the decay length is between 100 $\mu$m
and a few m.
In the case of $\na \to \gamma + G$, future detectors 
will be able resolve displaced $\gamma$ tracks at the level of a few cm. 
Alternately, the rarer decay mode $\na \to Z^0 + G$ gives rise to 
displaced charged particle vertices, which can be measured down to the
100 $\mu$m level. 

If $\na$ is mostly Higgsino, it decays (in the decoupling limit) predominantly
by $\na \to h^0 + G$.  
In the mostly Higgsino region of parameter space the states 
$\na$, $\capos$, and $\nb$, $\caneg$ form approximate $SU(2)_L$ doublets
with splittings much smaller than the overall mass scale. 
At both $e^+e^-$ and hadron colliders the signatures are therefore 
$h^0 h^0 X + \EmissT$ with $h^0 \to bb$,
where $X$ represents off-shell
$W^*$ and $Z^*$ electro-weak cascades of the heavier Higgsino states to the
lightest one. 
The existence of 4 $b$-jets which reconstruct $m_{h^0}$ in pairs, 
along with missing energy is therefore
also a distinctive feature of low scale supersymmetry breaking. 
In this case, displaced $b$-jet vertices could be measured down to the 
100 $\mu$m level.
Finally, if $m_{H^0}, m_{A^0} < m_{\na}$ the decays $\na \to H^0 +G$, and
$\na \to A^0 + G$,
with $H^0$ and $A^0$ undergoing standard model like decays,
would represent a gold-mine for Higgs physics. 

\subsection{Slepton NLSP}

With low scale supersymmetry breaking it is equally possible that
the lightest standard model superpartner is a charged slepton. 
For example, a gauge-mediated 
messenger sector with two generations of ${\bf 5} + \bar{\bf 5}$ 
generally gives a right handed slepton as the NLSP. 
In this case it decays by $\tilde{l}_R \to l + G$. 
These decays also lead to very distinctive signatures. 

At an $e^+e^-$ collider the signature for slepton pair production 
with prompt decay to Goldstinos is $e^+e^- \to l^+l^- + \Emiss$. 
The standard model backgrounds are easily manageable for this signal. 
The decay leads to a flat lepton spectrum in the lab frame,
with the end points giving an important test that the missing
energy is carried by essentially massless particles. 

At a hadron collider pair production of NLSP sleptons
with prompt decay to Goldstinos also gives final states
$ l^+l^- + \EmissT$. 
This suffers from fairly large irreducible backgrounds,
and does not represent a clean signature. 
However, production of heavier states which cascade to 
$\tilde{l}_R$ can give clean signatures with multiple leptons. 
For example, pair production of $\lL \lL$ 
followed by the cascade decays $\lL^{\pm} \to \tilde{l}_R^+ l^- l^{\pm}$
(through on- or off-shell $\chi_i^0$)
gives final states $6l + \EmissT$.
Such signatures do not suffer contamination by standard model backgrounds. 

If the decay $\tilde{l} \to l + G$ takes place on the scale of 
a detector very distinctive charged particle tracks can arise. 
This is because heavy non-relativistic sleptons are more highly ionizing
than ultra-relativistic charged particles. 
Decay over a finite distance can then be seen as a greater than minimum
ionizing track with a kink to minimum ionizing track. 
Such kinks should be measurable down to the 100 $\mu$m level. 
If the slepton decay length is long enough, 
timing information can also be applied to isolate such events. 
Measurement of the decay length distribution would give a direct
measure of the supersymmetry breaking scale. 

Because of the larger Yukawa coupling, the $\tilde{\tau}_R$ can be 
lighter than $\tilde{\mu}_R$ and $\tilde{e}_R$ from renormalization group
evolution. 
If $m_{\tilde{\tau}_R} + m_{\tau} + m_{\mu} < m_{\tilde{\mu}}$
the electro-weak decay 
$\tilde{\mu}_R^{\pm} \to \tilde{\tau}_R^+ \tau^- \mu^{\pm}$, 
through the $B$-ino component of off-shell $\chi_i^0$,
can compete with the decay $\tilde{\mu} \to \mu + G$,
and likewise for $\tilde{e}_R$. 
It is possible then that 
nearly all cascades lead to $\tilde{\tau}_R$ and that all 
the slepton signatures discussed above
occur with $\tau$'s in the final state. 
Alternately,
if $m_{\tilde{\tau}_R} + m_{\tau} + m_{\mu,e} > m_{\tilde{\mu},\tilde{e}}$,
all the right handed sleptons
are stable against three body electro-weak decays
at lowest order, and the decay $\tilde{l}_R \to l + G$
can dominate.  
In this case the above signatures occur with equal rates for all three
generations.

If the supersymmetry breaking scale is well above a few 1000 TeV,
the decay $\tilde{l} \to l + G$ takes place well outside the 
detector. 
At both $e^+e^-$ and hadron colliders the signature for supersymmetry
would then be $\lR^+ \lR^- X$, i.e. heavy charged particle pair production
without missing energy!  
This very non-standard signature should not be overlooked in 
the search for low scale supersymmetry breaking.

\subsection{Model Independent Signatures}

As discussed above,
the quantum numbers and composition of the lightest standard model
superpartner (NLSP) are model dependent. 
Although attention in the literature has been focused on the 
the mostly gaugino neutralino case, it is important to allow for
all possibilities in the search for low scale supersymmetry breaking, and 
decays to the Goldstino. 
From the above discussion, the model independent signatures break up into 
two exclusive possibilities
\begin{enumerate}
\item Neutralino NLSP
 \begin{itemize}
   \item $\gag X + \EmissT$
   \item $\gamma bb  X + \EmissT$
   \item $bbbb X + \EmissT$
 \end{itemize}
\item Slepton NLSP
  \begin{itemize}
  \item Multi-leptons + $\EmissT$, or 
  \item Heavy charged particle pairs
  \end{itemize}
\end{enumerate}
The precise form of $X$ results from cascade decays and can yield interesting
information about the superpartner spectrum and composition of low lying
states. 
$X$ often contains leptons, which helps to reduce standard model backgrounds.

\section{Conclusions}

If the messenger scale for supersymmetry breaking is not too far
above the weak scale the existence of the essentially massless Goldstino
can lead to very distinctive experimental signatures. 
For a neutralino NLSP final states with $\gag$, $\gamma bb$, or 
$bbbb$ and missing energy represent an important signature for decay
to the Goldstino. 
For a slepton NLSP final states with multi-leptons and missing energy
or heavy charged particles with kinks to minimum ionizing tracks
can result from decay to the Goldstino. 
A measurement of the decay length distribution to the Goldstino
would give a measurement of the supersymmetry breaking scale. 
Supersymmetry breaking at a low scale clearly provides many 
experimental challenges and opportunities. 

We would like to thank A. Litke and A. Seiden for useful discussions. 
This work was supported by the Department of Energy (M.D.),
under contract DOE-ER-01545-646 (S.R.), and
DOE-AC03-76SF00515 (S.T. and J.W.).


\begin{thebibliography}{9}

\bibitem{lsgauge}
M. Dine, W. Fischler, and M. Srednicki, Nucl. Phys. B
{\bf 189} (1981) 575;
S. Dimopoulos and S. Raby, Nucl. Phys. B {\bf 192} (1981) 353;
M. Dine and W. Fischler, Phys. Lett. B {\bf 110} (1982) 227;
M. Dine and M. Srednicki, Nucl. Phys. B {\bf 202} (1982) 238;
L. Alvarez-Gaum\'{e}, M. Claudson, and M. Wise, Nucl. Phys. B
{\bf 207} (1982) 96;
C. Nappi and B. Ovrut, Phys. Lett. B {\bf 113} (1982) 175.

\bibitem{hsgauge} M. Dine and W. Fischler, Nucl. Phys. B {\bf 204} (1982) 346;
S. Dimopoulos and S. Raby, Nucl. Phys. B {\bf 219} (1983) 479.

\bibitem{fayet} P. Fayet, Phys. Lett. B {\bf 70} (1977) 461;
P. Fayet, Phys. Lett. B {\bf 84} (1979) 416;
R. Casalbuoni, S. De Curtis, D. Dominici, F. Feruglio,
and R. Gato, Phys. Lett. B {\bf 215} 313.

\bibitem{ddrt} M. Dine, S. Dimopoulos, S. Raby, and S. Thomas,
hep-ph/9601367, Phys. Rev. Lett. {\bf 76} (1996) 3484.

\bibitem{dn} M. Dine, A.E. Nelson and Y. Shirman, 
Phys. Rev. D {\bf 51} (1995) 1362;
M. Dine, A.E. Nelson, Y. Nir and Y. Shirman,
Phys. Rev. D {\bf 53} (1996) 2658.

\bibitem{stump} D. Stump, M. Weist, and C. P. Yuan, hep-ph/9601362.

\bibitem{dtw} S. Dimopoulos, S. Thomas, and J. D. Wells, 
hep-ph/9604452, to appear in Phys. Rev. D. 

\bibitem{gordy} S. Ambrosanio, G. Kane, G. Kribs, and S. Martin,
hep-ph/9605398.


\end{thebibliography}
\end{document}